\documentclass{IEEEmce}

\usepackage[colorlinks,urlcolor=blue,linkcolor=blue,citecolor=blue]{hyperref}
\usepackage{upmath}
\jvol{XX}
\jnum{XX}
\paper{8}
\jmonth{xxx/xxx}
\publisheddate{00 xxxx 0000}
\currentdate{00 xxxx 0000}
\jname{IEEE Software Magazine}
\pubyear{2023}
\doiinfo{Doi Number}

\usepackage[acronym, hyperfirst=false]{glossaries}

\newcommand{\rqoneOccur}{RQ1: What experiences lead to psychological overwhelm within software development activities?}
\newcommand{\rqoneOccurShort}{What led to psychological overwhelm?}
\newcommand{\rqtwoExperience}{RQ2: How do software developers experience overwhelm in their daily work?}
\newcommand{\rqtwoExperienceShort}{How was overwhelm experienced?}
\newcommand{\rqthreeProductivity}{RQ3: How do software developers experience overwhelm in relation to their productivity?}
\newcommand{\rqthreeProductivityShort}{How was productivity affected?}
\newcommand{\rqfourStress}{RQ4: What role does stress play in developer's experience of overwhelm?}
\newcommand{\rqfourStressShort}{What role does stress play?}

\newacronym{ipa}{IPA}{Interpretive Phenomenological Analysis}

\usepackage{amsmath}

\begin{document}

\sptitle{Department: Head}
\editor{Editor: Name, xxxx@email}

\title{Overwhelmed Software Developers}
\author{\,Lisa-Marie Michels}
\affil{University of Stuttgart, Germany}
\author{Aleksandra Petkova}
\affil{University of Stuttgart, Germany}
\author{Marcel Richter}
\affil{University of Stuttgart, Germany}
\author{Andreas Farley}
\affil{University of Stuttgart, Germany}
\author{Daniel Graziotin}
\affil{University of Hohenheim, Germany}
\author{Stefan Wagner}
\affil{Technical University of Munich, Heilbronn, Germany}
\markboth{Department Head}{Paper title}

\begin{abstract}
We have conducted a qualitative psychology study to explore the experience of feeling overwhelmed in the realm of software development. Through the candid confessions of two participants who have recently faced overwhelming challenges, we have identified seven distinct categories: communication-induced, disturbance-related, organizational, variety, technical, temporal, and positive overwhelm. While most types of overwhelm tend to deteriorate productivity and increase stress levels, developers sometimes perceive overwhelm as a catalyst for heightened focus, self-motivation, and productivity. Stress was often found to be a common companion of overwhelm. Our findings align with previous studies conducted in diverse disciplines. However, we believe that software developers possess unique traits that may enable them to navigate through the storm of overwhelm more effectively.
\end{abstract}

\maketitle

\enlargethispage{10pt}

\chapterinitial{Emotion}, a trait that sets humans apart from machines, can be both a powerful tool and a potential hindrance. When harnessed positively, emotions can foster creativity and enhance efficiency \cite{TheRoleofAffectiveExperience}. In particular, positive emotions play a crucial role in boosting productivity in software engineering \cite{graziotin2018What,graziotin2015How}.

Furthermore, negative emotions can have detrimental effects on code quality, cognitive performance, and overall productivity. They can lead to work withdrawal and a general decrease in efficiency \cite{graziotin2018What}. While negative emotions, for example, conflict-induced anger may ignite problem-solving abilities, there consensus within the software engineering discipline that minimizing negative emotional experiences is crucial \cite{graziotin2018What} and, we add, moral and ethical.

A common outcome of negative emotions is the experience of ``psychological overwhelm,'' a condition that includes a range of emotions from occasional doubt to chronic despair \cite{lipskyAgeOverwhelmedStrategies2018}. In this paper, we adopt the noun form of ``overwhelm'' to refer to ``the act of overwhelming, or the fact of being overwhelmed'' as defined by the Oxford English Dictionary. 

Overwhelm is a universal phenomenon, yet academia knows so little about it \cite{FeelingOverwhelmedAParsesciencingInquiry}, and it remains an ambiguous concept that has not agreed-upon definition. \cite{FeelingOverwhelmedAParsesciencingInquiry,doi:10.1177/0894318418807931}. Still, we all relate to overwhelm in our everyday tasks, including those of software engineers. These professionals are often subject to high-stress levels \cite{ostberg2020methodology,akula2008impact}. The connection between software engineering, overwhelm, and stress is still uncertain.

Our objective is to bridge these gaps by exploring the experiences of developers when they encounter overwhelm, examining its impact on their productivity, and investigating the role of stress in this process \cite{michels2022automatic}.

\glsresetall
We are inspired by Lenberg et al.'s recent call in the realm of behavioral software engineering \cite{lenberg2023qualbse}. They advocate for the use of a wider range of qualitative methods from the social and behavioral sciences, emphasizing the importance of researchers' reflexivity. We have opted to utilize \gls{ipa}, a qualitative research approach not yet prevalent in software engineering, but fitting for our objective \cite{lenberg2023qualbse}.

Psychological overwhelm often results from a sequence of experiences culminating in an overwhelmed state˜\cite{doi:10.1177/0894318418807931}. This observation forms the basis of our first line of research, which involves dissecting the experiences that trigger psychological overwhelm. Our first research question is:

\textbf{\rqoneOccur} \
Building upon the previous discussion, our second research question focuses on exploring the nature of experiences encountered during states of psychological overwhelm:

\textbf{\rqtwoExperience}\
We also aim to investigate how psychological overwhelm impacts the productivity of software developers. Following the guidelines of the \glsentrytext{ipa}, we assess productivity based on personal accounts provided by our interviewees, leading us to our third research question:

\textbf{\rqthreeProductivity} \
Narratives often portray overwhelm when stressors exceed one's capacity, causing neural overload and potential physical and mental repercussions \cite{StressOverwhelm}. To gain an all-embracing understanding of the connection between stress and overwhelm, we analyze stress-related themes emerging from our \glsentrytext{ipa} interviews. Thus, our final research question:

\textbf{\rqfourStress} \

\noindent
\begin{figure}[h]
\small
\centering
\fbox{%
	\begin{minipage}{\dimexpr\columnwidth-2\fboxsep-2\fboxrule\relax}
	\glsresetall
	\textbf{\gls{ipa}} \\
	\glsentrytext{ipa} is a robust research method used in the social sciences to delve into individuals' understanding of their personal and social experiences. One of its distinguishing features is the deliberate selection of small, yet meaningful, sample sizes (even a single participant!), coupled with exploratory research questions aimed at eliciting rich insights.
	
	As a methodology, \glsentrytext{ipa} excels in uncovering the cognitive, emotional, and perceptual aspects of individual experiences. In the context of software engineering, \glsentrytext{ipa} holds great promise. It enables the exploration of subjective experiences of stakeholders, capturing their challenges and potential areas for improvement, and providing a human lens through which software development can be understood.

	In a recently published review article \cite{lenberg2023qualbse}, we have discussed \glsentrytext{ipa}, along with two other novel qualitative research methods (new to SE), reflective practices, and the COREQ quality criteria for qualitative studies, which we followed when we conducted the present study. Refer to it for a more comprehensive understanding of \glsentrytext{ipa} and new qualitative research practices for software engineering research.

	\end{minipage}%
}
\end{figure}

In the realm of software engineering, we are venturing into unexplored territory both in industry and academia. The subject we are delving into, the experience of overwhelm, has not been extensively analyzed \cite{FeelingOverwhelmedAParsesciencingInquiry}. Our paper offers initial insights into the topic of ``overwhelm'', which is relevant to daily work in the software industry and requires preventive measures in the future. Moreover, the utilization of \glsentrytext{ipa} for research in this context is groundbreaking. We are embarking on a journey to uncover correlations between overwhelm, stress, cognitive overload, and negative emotions.

Kabigting's research \cite{doi:10.1177/0894318418807931} synthesizes insights from various disciplines regarding overwhelm. They identify three crucial themes. Firstly, overwhelm is depicted as a cascading disaster that unexpectedly plunges people, leading to feelings of entrapment and suffocation. Secondly, overwhelm tends to amplify feelings of solitude and helplessness. Lastly, people often resort to disruptive or harmful means to cope with the intense emotions.

There is also a limited body of work examining the stress faced by software engineers. Professionals are aware of the inherent stress in their roles \cite{akula2008impact,ostberg2020methodology} and openly discuss its downsides \cite{graziotin2018What}. Assessing stress of software engineers is a complex task, but we are making progress towards effective measurement techniques \cite{ostberg2020methodology}.

Iskander \cite{iskander_burnout_2019} discusses cognitive overload as the point at which working memory becomes overwhelmed, leading to overflow into long-term memory. Their research reveals that symptoms of cognitive overload, such as increased error rates and declines in habitual competence, intersect with symptoms of burnout. Therefore, monitoring cognitive overload can serve as an early warning sign of potential burnout.

The wide range of negative emotions associated with overwhelm motivates us to evaluate their impact on developer productivity. Our previous research has indeed touched upon the interplay of positive and negative experiences in software development \cite{graziotin2015How,graziotin2018What}.

\section{Methodology}\label{sec:Methodology}
\glsresetall

We adopted \gls{ipa} \cite{InterpretativePhenomenologicalAnalysis} as our research framework. \glsentrytext{ipa} focuses on exploring people's experiences of a particular phenomenon. This approach necessitates a thorough examination of individual cases, resulting in a relatively small sample size. \glsentrytext{ipa} research does not aim to test hypotheses, but rather seeks to understand the area of interest. The application of \glsentrytext{ipa} in the context of software engineering was introduced by Lenberg et al. \cite{lenberg2023qualbse}. Refer to our sidebar named \textbf{Interpretive Phenomenological Analysis (IPA)} to learn more about it.

\subsection{Data Collection}

Participants for this study were recruited among professional peers following state-of-art ethical principles. We conducted semi-structured interviews for data collection, a technique frequently used in \glsentrytext{ipa} studies \cite{brocki2006critical}. Despite the limited guidance for designing effective interviews in many \glsentrytext{ipa} studies \cite{brocki2006critical}, we created an interview guide with prompt questions. This guide is available in an online technical report (refer to \textit{Further Reading}). The interviews were recorded and transcribed, and these artifacts were deleted post-analysis following ethical guidelines and privacy laws.

\subsection{Data Analysis}

The analysis itself was divided into four stages, as proposed by Smith et al. \cite{smith2002risk}. In the first stage, all team members read the first transcript and made brief notes without assigning any labels or themes. The purpose of this stage was to gain an overview of the transcript. In the second stage, labels were created and assigned to the relevant statements. After completing the second stage, team members exchanged their labels and reviewed them. This led to the third stage, where team members collectively clustered the themes and assigned names to them. In cases where contrasting perspectives emerged, team members voted on how to proceed. In the final and fourth stage, the synchronized clusters were aggregated and summarized.

\section{Results}

We present the results of the study on a case-by-case basis. Each participant is associated with emerging themes following Wiling's  \cite{wiling2008introducing} work. The curated themes represent a combined perspective of each author's analysis.

Two individuals participated in the study, referred to by the fictional names James and Charles. James is 44 years old, while Charles is 27 years old. Both identify as males and work within IT-based companies as software developers, without any managerial roles.

Through our analysis, we identified seven types of overwhelm, which we refer to throughout the text and identify in the box named \textbf{The distinct experienced overwhelm}.

\begin{figure}[h]
\small
\centering
\noindent
\fbox{%
	\begin{minipage}{\dimexpr\columnwidth-2\fboxsep-2\fboxrule\relax}
	\textbf{The distinct experienced overwhelm} \\
	\textit{Communication Overwhelm}, where developers are overwhelmed when handling critical social situations.\\
	\textit{Disturbance Overwhelm}, where developers are overwhelmed by interruptions from meetings, colleagues, reassignment, or noise.\\
	\textit{Organizational Overwhelm}, where developers are overwhelmed by the necessity to prioritize and organize their workloads.\\
	\textit{Variety Overwhelm}, where developers are overwhelmed by the sheer number of tasks.\\
	\textit{Technical Overwhelm}, where developers are overwhelmed by the complexity or magnitude of individual tasks.\\
	\textit{Temporal Overwhelm}, which includes feelings of being overwhelmed due to time constraints.\\
	\textit{Positive Overwhelm}, where developers perceive overwhelm positively, driving their ambition.
	\end{minipage}%
}
\end{figure}

\subsection{James' Overwhelm}

James, a 44-year-old developer, works in a hybrid role, dividing his time between coding, software testing, and mentorship within his occupational field. His workload allocation roughly comprises 10\% integrating new lines of code, 10\% merging externally written code, 30\% code testing, with the remaining portion dedicated to training and mentorship. Despite starting without a formal degree, he eventually pursued higher education after acquiring considerable industry experience.

\subsubsection{\rqoneOccurShort} 

Before the interview, James had categorized his experiences of overwhelm into two pillars: \textit{Temporal Overwhelm} and \textit{Technical Overwhelm}. For him, \textit{Temporal Overwhelm} arises from many small problems compressed within limited timeframes or technical challenges that demand more time than he can afford (``Sometimes we ran to keep the machines [\ldots] alive''). The equation of \textit{Technical Overwhelm} to \textit{Temporal Overwhelm} suggests that task difficulty is bearable if adequate time is available for problem-solving. Dealing with old legacy code, particularly when the original authors were no longer with the company, was a typical technical task that overwhelmed him.

Additionally, we identified two main sources of \textit{Temporal Overwhelm} for James: workload quantity and the respective time required for completion. In extreme cases, \textit{Temporal Overwhelm} escalated when he worked ten-hour days—the maximum permitted in Germany—sometimes continuing for five years.

Solitude in his professional world was highly valued, as mental focus is fundamental to James' work. Distractions, such as demands for consultation from customers or untimely requests from superiors, derailed his focus and led to overwhelm.

Another source of overwhelm was the pressure from role transitions, such as when James temporarily moved into a managerial role and became a liaison between customers and developers. This position made him the center of everyone's attention and increased his stress levels. He attributed this experience to his decision never to accept a managerial role.

\subsubsection{\rqtwoExperienceShort} 

James associates the experience of overwhelm with individual personality traits and developers' experience levels, which also determine their reactions to different sources of overwhelm.

In his early career, James had a naive mentality driven by a do-or-die mindset accompanied by adrenaline, which led him into constant overwhelm, especially on big projects. Over time, physiological symptoms like hair loss due to stress manifested, and psychological symptoms like lack of focus emerged (``I become crippled, and I cannot focus properly anymore''). Project burnouts were quite common, with four out of fifty programmers experiencing them in one project he recollected.

To prevent further harm, James consciously decided to slow down and developed coping strategies such as limiting his daily tasks to three and fending off distractions. He attributed his improved task estimation skills to his experience, which played a crucial role in managing \textit{Temporal Overwhelm}.

As he matured, his sense of responsibility lessened, and he learned to shrug off pressures. He underlined the role of superiors in shielding engineers from unreasonable requests, especially during \textit{Temporal Overwhelm}. Over time, the emotional response to \textit{Temporal Overwhelm} shifted from dread to frustration and sadness.

\textit{Disturbance Overwhelm}, experienced as intense anger and frustration, surfaces when James is thrown out of his work zone by distractions or task reassignment (``A customer calls and that, for me, is stressful''). Dealing with such situations became easier as he grew more confident in standing his ground against superiors and co-workers. Despite irritation due to colleague interruptions, James considered it necessary since he also depended on his colleagues' assistance at times.

To minimize disturbances, James isolated himself, communicated sparingly, and used noise cancellation devices. Despite attempting to confine work within office hours from day one, he often found himself pondering about his pending tasks.

\subsubsection{\rqthreeProductivityShort} 

Interestingly, James noted that his productivity occasionally improved during instances of \textit{Temporal Overwhelm}, attributing it to the adrenaline surge that helped him ``get into the zone.'' Conversely, he experienced significant productivity loss during episodes of \textit{Disturbance Overwhelm} as distractions interrupted his workflow, necessitating frequent resets of focus. During such periods, his irritability occasionally manifested as rudeness towards other team members.

\subsubsection{\rqfourStressShort} 

Unprompted, James openly acknowledged stress as a significant element in his work life, particularly during times of overwhelm. He revealed that he would intentionally induce stress as a coping mechanism for dealing with \textit{Temporal Overwhelm}. In the case of \textit{Disturbance Overwhelm}, stress was often associated with customer calls and internal communication tools such as Skype and Teams. This form of pressure would disrupt his internal calm and divert his focus away from productive work (``I’d not say that one starts shaking, but I did feel an inner shaking'').

During his previous managerial roles, James experienced what he referred to as ``psychic stress,'' which was primarily triggered by negative comments from others. However, he noted that this experience was not shared in his current position. Additionally, stress would manifest in his tendency to continuously replay work scenarios and engage in problem-solving even during his off-work hours.

\subsection{Charles' Overwhelm}

Charles, a 27-year-old IT Consultant specializing in advising banks on technical issues, shared his experiences. His primary responsibility involves understanding the technical challenges faced by banks, gathering relevant data, and utilizing low-code development platforms to devise suitable solutions. Unlike James, Charles operates at a higher level of abstraction, with his work rarely involving conventional programming. However, he does create software for testing and validating his proposed solutions. Charles holds a Master's degree in Business Information Technologies and began working immediately after completing his studies.

\subsubsection{\rqoneOccurShort} 

The interview with Charles revealed multiple instances of overwhelm in his work life. One such instance was \textit{Temporal Overwhelm}, which arose when Charles faced a high volume of small tasks and tight deadlines. He was ``overwhelmed from the sheer mass''. This situation often resulted in longer working hours and increased work pressure. The tasks themselves also contributed to feeling overwhelmed, as Charles is only ``working with bad work'' (meaning, bad architectural choices or low-quality code). This includes tasks Charles considers as ``just crap'', and tasks which are ``documented badly or incorrectly''. Temporal Overwhelm was also magnified when getting close to the aforementioned deadlines, by superiors asking if it is ``necessary that people work on the weekends'', giving Charles the feeling of ``okay, this is looking really bad''. 

Additionally, Charles experienced \textit{Organizational Overwhelm} when struggling to organize and prioritize his tasks. The lack of proper task documentation further compounded this challenge, leading to confusion about where to begin.

Seeking assistance from other departments or customers also caused overwhelm for Charles, particularly during his initial experiences. He was worried about making mistakes or saying bad things, which made him very stressed and anxious.

Charles experienced \textit{Disturbance Overwhelm} due to regular intrusions from his colleagues while working on complex problems. His preference is to focus on individual, significant problems, deriving fulfillment from resolving them. However, he is agitated when people persistently engage him for unrelated topics throughout the day. The constant communication and interruptions from his peers overwhelm him. 

Charles recalled a situation when the Austrian branch of his company was not on holiday while the German branch was. The Austrians reported exceptional productivity the next day, emphasising the lack of disruptions. This led to internal management discussions about possibly limiting interruptions among peers. He appreciated the benefits of support from seasoned peers, understanding the fine equilibrium it offered.

Lastly, Charles identified instances of \textit{Positive Overwhelm} where a substantial workload actually helped him maintain a sharper focus.

\subsubsection{\rqtwoExperienceShort} 

Charles experienced \textit{Positive Overwhelm} as a form of pressure that heightened his work focus. However, \textit{Technical Overwhelm} and \textit{Disturbance Overwhelm} elicited negative emotions such as annoyance, anger, and confusion. Charles often described tasks which were, in his opinion, of poor quality, and working on such tasks could trigger a wide range of emotions, including being ``annoyed to death'' by strenuous tasks.
Poor code quality and inadequate documentation contributed to feelings of helplessness and self-doubt. Struggling with problems made him ``feel [\dots] overwhelmed because you are thinking, you are just too stupid''.

\textit{Communication Overwhelm} triggered anxiety about the possibility of misspeaking (anxiety to ``say the wrong thing''), while \textit{Organizational Overwhelm} resulted in confusion and depleted energy. To cope with overwhelm, Charles learned to seek help from others, accept his limited knowledge, improve task prioritization, and distance himself from undue responsibilities.

The mitigation of overwhelm symptoms primarily relied on effective management. Sound leadership that shielded the team from unnecessary external pressure, provided clear direction, and prioritized tasks facilitated a sense of freedom and increased productivity. Easy access to experienced colleagues and a cooperative work environment also helped alleviate feelings of overwhelm.

Despite experiencing stress, Charles did not report any physiological symptoms. However, he often carried work-related thoughts home with him.

\subsubsection{\rqthreeProductivityShort} 

Charles experienced overwhelm when faced with excessive upfront information on a new project, which impaired his productivity (There is ``a lot of information you need to know, to work productively''). Constant distractions also had a significant impact on his productivity by disrupting his focus. However, in instances of \textit{Positive Overwhelm}, Charles reported heightened productivity.

\subsubsection{\rqfourStressShort} 

Charles reported experiencing stress both as a positive force that enhanced his productivity and as a negative by-product of overwhelm. Interestingly, he confirmed the correlation between feelings of stress and overwhelm without direct prompting from the interviewer. This suggests that Charles recognized the complex relationship between stress and overwhelm in his work life.

\begin{figure}[h]
\small
\centering
\noindent
\fbox{%
	\begin{minipage}{\dimexpr\columnwidth-2\fboxsep-2\fboxrule\relax}
	\textbf{Overwhelm, Stress, and Productivity: An Overview of James and Charles' Experiences} \\
	\paragraph{Occurrence of Overwhelm} Both James and Charles attributed the feeling of overwhelm to various sources, including code quality issues and frequent task reassignments. Additionally, they both noted that heavy workloads and time crunches contributed to overwhelm. Both participants also observed that distractions amplified their feelings of overwhelm.
	
	\paragraph{Experience of Overwhelm} Both participants reported negative experiences with \textit{Temporal Overwhelm} and \textit{Technical Overwhelm}, which were the only two types of overwhelm common to both interviews. While they mentioned other types of overwhelming situations, only Charles reported some encounters that resulted in positive outcomes. Frustration was a common emotion associated with overwhelm for both participants. James also reported experiencing physiological symptoms such as hair loss, excessive tiredness, and concentration problems.
	
	\paragraph{Impact on Productivity}
	
	Both participants acknowledged that overwhelm could have both negative and positive impacts on their productivity levels. For example, \textit{Temporal Overwhelm} could lead to heightened focus and positive productivity outcomes, while \textit{Technical Overwhelm}, often associated with juggling multiple tasks, had a detrimental effect on productivity.
	
	\paragraph{Impact on Stress}
	
	Both participants spontaneously mentioned stress, indicating a close connection between stress and feelings of overwhelm. The stress discussed by the participants was often self-imposed and seen as a negative factor in James's situation. However, Charles acknowledged that stress occasionally had positive impacts on his productivity.
	\end{minipage}%
}
\end{figure}

\section{Discussion}\label{sec:Discussion}
In this section, we will discuss our findings related to overwhelm (RQ1 and RQ2), productivity (RQ3), and stress (RQ4).

\subsection{Overwhelm}
Our primary objective was to explore the experiences of developers when they feel overwhelmed, and we believe that our study successfully achieved this goal. We identified several themes of overwhelm, including communication-induced, disturbance-related, organizational, variety, technical, temporal, and positive overwhelm. In a literature review conducted by Kabigting \cite{doi:10.1177/0894318418807931}, three themes of overwhelm were formalized. Kabigting \cite{doi:10.1177/0894318418807931} describes a theme of overwhelm associated with sudden engulfment, often accompanied by feelings of being trapped or drowned. While our participants shared similar experiences, with James likening overwhelming stress to a ``crippling'' sensation, they did not recount a sudden onset of overwhelm. Instead, they pointed to persistently high-pressure leading to stress and eventually overwhelm.

The theme of overwhelm associated with feelings of isolation or powerlessness did not explicitly emerge from our interviews, although there were some indications. Charles described a sense of helplessness when overwhelmed, but did not report feeling isolated.

Finally, the theme of coping mechanisms for overwhelm includes reaching out to others for assistance, lashing out, or engaging in self-harm. Components of these coping strategies were reflected in our interviews, as both participants reached out to colleagues or superiors to mitigate their overwhelming scenarios.

\subsection{Productivity}
Before conducting the interviews, we anticipated that participants would report a loss of productivity associated with overwhelm. However, we also discovered instances where overwhelm was linked to gains in productivity. Both participants discussed how self-imposed pressure, often resulting from overwhelm, could lead to heightened focus and enhanced productivity. However, this could come with its risks, such as burnout.

\subsection{Stress}
Stress was frequently mentioned by the participants. They struggled to precisely define what stress meant to them, often using the terms ``pressure'' and ``stress'' interchangeably. The prevalence of stress mentions may indicate the participants' challenge in identifying the specific emotions they experienced during overwhelm, resorting to the term stress, which can encompass various emotional strains.

\section{Conclusion}\label{sec:Conclusion}
\glsresetall
We explored the experiences of software developers when overwhelmed, employing an \gls{ipa} approach for rich, detailed insights into their feelings, experiences, and the impact on their work.

Seven themes of overwhelm surfaced in our study: communication, disturbance, organizational, variety, technical, temporal, and positive overwhelm. Each theme posed unique challenges and was associated with emotions from distress to ambition. Temporal and technical overwhelm were identified as particularly challenging by the participants.

Stress featured prominently in participants' discussions, with terms like ``pressure'' and ``stress'' frequently used to describe their experiences. This underlines the interconnectedness of these experiences, often linked to sustained high workloads and persistent, sometimes self-imposed, demands.

Our participants identified unique characteristics that distinguish our findings from other domains. These insights could assist software development teams, managers, and organizations in formulating strategies to effectively manage overwhelm and stress among developers. While our study did not focus on remediation strategies, the participants offered several suggestions. Creating non-competitive workplace cultures and shielding employees from external pressures are seen as effective measures to reduce overwhelm.

Additionally, management should proactively tackle the issue of overwhelm by employing specific strategies. These include appropriately planning tasks, providing information in manageable doses, and aligning the actual work with the resources and capabilities of individual employees. Implementing regular breaks, specific workload management techniques, and psychological wellbeing practices at work are also recommended to mitigate stress and overwhelm.

Our participants mentioned instances where their colleagues or superiors, who had personal stakes in the projects, exhibited notable physiological and psychological reactions to overwhelming situations, such as extreme fatigue, sleep difficulties, and burnout. An intriguing avenue for future investigation involves exploring the experiences of team leaders or senior developers who find themselves overwhelmed, particularly those who have personal stakes in the company.

Including participants who have faced severe physiological or psychological responses to overwhelm, like extreme fatigue, insomnia, or burnout, could deepen our comprehension. Although less common, the effects of positive overwhelm present an intriguing research opportunity. To gain more profound insights, researchers could consider recruiting individuals who have experienced these symptoms. It appears that personality traits play a significant role in how individuals perceive and handle overwhelm, with some individuals being more susceptible to its negative effects. Interestingly, our participants demonstrated a remarkable ability to effectively manage overwhelming situations.

The robust connection between overwhelm and stress, as evidenced in this study, necessitates further exploration of this relationship's nature and dynamics within the software engineering context.

\section{Further reading}\label{sec:further-reading}
The full technical report is openly and freely available at \url{https://arxiv.org/abs/2401.02780}. In it, we provide the full details regarding the research design including reflective practices, the interview guide, the data analysis method, and the results, which are all substantiated by interview quotes and summary tables. Additionally, we offer a more elaborate comparison between the experiences of James and Charles.

\subsection{Acknowledgment}
We are grateful to James and Charles for sharing their experiences with us.

\bibliographystyle{IEEEtran}
\bibliography{bibi}

\begin{IEEEbiography}{Lisa-Marie Michels}{\,} is with the Institute of Software Engineering, University of Stuttgart, Germany. She is a former PhD student in computer science from the University of Stuttgart, Germany. Her research focuses on empirical studies on mental wellbeing and affective computing, using a combination of qualitative and quantitative research methods from psychology and neuroscience.
\end{IEEEbiography}

\begin{IEEEbiography}{Aleksandra Petkova}{\,} is a student at the University of Stuttgart, Germany.
\end{IEEEbiography}

\begin{IEEEbiography}{Marcel Richter}{\,} is a student at the University of Stuttgart, Germany.
\end{IEEEbiography}

\begin{IEEEbiography}{Andreas Farley}{\,} is a student at the University of Stuttgart, Germany.
\end{IEEEbiography}

\begin{IEEEbiography}{Daniel Graziotin}{\,} is a full professor of information systems and digital technologies at the University of Hohenheim, Germany. He earned his PhD in computer science and software engineering from the Free University of Bozen-Bolzano, Italy. His research focuses on interdisciplinary and multidisciplinary approaches, incorporating theories, methods, and measurements from social and behavioral sciences, to enhance the understanding and integration of human factors in technology development and implementation.
\end{IEEEbiography}

\begin{IEEEbiography}{Stefan Wagner}{\space}(Senior Member, IEEE) is a full professor of  software engineering at the Technical University of Munich, Heilbronn, Germany. He studied psychology in Hagen
and computer science in Augsburg and Edinburgh, and he holds a doctoral degree
in computer science from the Technical University of Munich. His research interests are empirical studies, software quality, human aspects, automotive software, and AI-based software. He is a Senior Member of ACM.
\end{IEEEbiography}

\end{document}